# Mechanism of Electricacupuncture Treating Detrusor–Bladder Neck Dyscoordination After Suprasacral Spinal Cord Injury by Proteomics

Li-Ya Tang[1†], Qi-Rui Qu[1†], Yi-Ying Long[1], Xia Wu[1], Jin-Can Liu[1], Ming Xu[1], Hong Zhang[1], Lu Zhou[2], Qiong Liu[1*], Kun Ai[1*].
1 College of Acupuncture, Massage and Rehabilitation, Hunan University of Chinese Medicine, Changsha, Hunan 410208, China.
2 Department of rehabilitation medicine, Chenzhou first people's Hospital, Chenzhou, Hunan 423000, China.
[†]The author contributes to this paper.
[*]Co-Corresponding authors:
Kun Ai, College of Acupuncture and Massage, Hunan University of Chinese Medicine, Changsha, Hunan 410208, China. Email: aikun650@qq.com.
Qiong Liu, College of Acupuncture and Massage, Hunan University of Chinese Medicine, Changsha, Hunan 410208, China. Email: 121181998@qq.com.

**Abstract**

**Objectives** This study aimed to elucidate the potential mechanisms of electroacupuncture (EA) in restoring detrusor-bladder neck dyssynergesia (DBND) following suprasacral spinal cord injury.

**Methods** A total of 52 adult female Sprague-Dawley rats were randomly assigned to either a sham group (n=12) or a spinal cord injury model group (n=40). In the model group, DBND was induced in 40 rats through Hassan Shaker spinal cord transection, with 24 rats surviving spinal shock and subsequently randomized into two groups: a model-only group (DBND, n=12) and an EA intervention group (DBND+EA, n=12). DBND+EA was administered at Ciliao (BL32), Zhongji (RN3), and Sanyinjiao (SP6) acupoints, for 20 minutes per session, once daily for 10 consecutive days. On day 29 post-injury, all rats underwent urodynamic assessments, followed by hematoxylin and eosin (HE) staining, tandem mass tag (TMT) proteomics, and Western blot (WB) analysis of the detrusor and bladder neck tissues.

**Results** Urodynamic evaluation demonstrated that EA intervention enhanced bladder function in DBND rats. HE staining indicated reduced fibroplasia in the detrusor muscle and alleviated inflammation in the bladder neck following EA. TMT proteomic analysis revealed 30 differentially expressed proteins (DEPs) in the detrusor and 59 DEPs in the bladder neck post-EA treatment. WB results corroborated these TMT findings.

**Conclusion** EA effectively promotes synergy between the detrusor muscle and bladder neck in DBND, likely by enhancing detrusor contractility and facilitating bladder neck relaxation during urination. This study provides mechanistic insights into the therapeutic role of EA in managing DBND.

## 1. Introduction

Neurogenic bladder is a usual side effect after suprasacral spinal cord injury (SSCI).

According to relevant statistics, 70-84% of patients with spinal cord injury (SCI) have bladder dysfunction [1]. The urinary function of the bladder is controlled by the bladder detrusor, bladder neck and musculus sphincter of external urethra. During normal bladder emptying, the parasympathetic nerve (S2-S4) in the spinal cord is activated, which cause bladder detrusor contracting. Furthermore, inhibition of the sympathetic nerves innervating the bladder neck (T11-L2) and the somatic nerves innervating the external urethral sphincter (controlled by Onuf's nucleus at S2-S4) results in bladder neck dilation, urethral relaxation, and urination. With SCI at T10 and above, the sympathetic and parasympathetic nerves innervating the bladder become disconnected from the pons and other higher urination centers, which results in autonomic dysreflexia (AD). In AD patients with SSCI, sympathetic overactivity leads to impaired bladder detrusor contractility and bladder neck overcontraction [2]. During urination, the detrusor and bladder neck are coordinated abnormally in a condition known as detrusor–bladder neck dyscoordination (DBND) [3]. If not treated in time, DBND may lead to urinary retention, urine reflux, kidney failure, and other serious urinary problems [4].

Therefore, DBND needs to be considered and addressed in the clinic. Alleviating excessive contraction of the bladder neck, promoting effective detrusor contraction, and applying dual-target interventions that address both issues should be the main clinical approaches used to improve DBND after SSCI. At present, most DBND treatment is targeted solely toward the bladder neck, and there are few reports on dual-target treatments addressing the bladder neck and detrusor. In clinical practice, DBND is mainly treated by surgery, intermittent catheterization and α-adrenergic blockers for the bladder neck, whereas there are few effective interventions for the bladder detrusor [5]. The current treatment methods have achieved some degree of clinical efficacy, but they tend to address single therapeutic targets, and they are accompanied by postoperative complications, drug side effects and other shortcomings.

Electroacupuncture (EA) is widely used to cure DBND because of its low invasiveness and lack of drug side effects, and its efficacy has been confirmed [6]. Studies have shown that acupuncture can bidirectionally regulate the function of diseased regions to restore a normal physiological state [7]. After patients with sphincter hyperactivity underwent acupuncture treatment, bladder output was increased, residual urine volume and bladder pressure decreased [8]. Additionally, in patients with detrusor hyperreflexia, acupuncture can increase the maximum bladder volume and reduce the frequency of urination [9]. These studies also suggest that acupuncture has a multitarget therapeutic effect, and the internal mechanism of acupuncture treatment for DBND likely involves the regulation of the detrusor and bladder neck. At present, the mechanism by which acupuncture treats neurogenic bladder mainly focuses on improving bladder detrusor function and external urethral sphincter or alleviating the loss of local spinal cord neurons [10]. In our study of EA as a treatment for DBND, we considered both studies observing the detrusor muscle and studies observing the bladder neck.

Proteins are the main molecules that carry out biological activities. Studies have shown that protein expression changes in spinal cord injury-induced neurogenic bladder [11]. Tandem mass tag (TMT) technology is a quantitative proteomics tool that can comprehensively detect proteins expressed in tissues. Bidirectional holistic regulation and multitarget effects are the main characteristics of acupuncture in the treatment of disease [12]. Therefore, the present study combines TMT quantitative proteomics with bioinformatics analysis and applies them to acupuncture research to explore the potential therapeutic targets and mechanisms of acupuncture in a holistic and multifaceted manner.

To date, TMT quantitative proteomics analysis has not been used to research the potential mechanism of EA treatmenting DBND after SSCI with detrusor muscle and bladder neck as dual therapeutic targets. In it, we transected T10 spinal cord segment to generating a DBND rat model and measured DEPs in the detrusor and bladder neck after EA treatment using TMT quantitative proteomics analysis. Bioinformatics analysis of DEPs was conducted to screen out key proteins and related signaling pathways that regulate detrusor and bladder neck contraction. In addition, we used Western blotting (WB) to detect these key proteins, which further elucidated the internal mechanism by which EA regulates detrusor and bladder neck contraction and improves urinary function in SSCI-induced DBND. This study provides a proteomic basis for EA treating DBND after SSCI.

## 2. Materials and methods

### 2.1 Animals origin and grouping

52 healthy adult female rats (weighing 250–280 g) were selected from the Laboratory Animal Center of Hunan University of Chinese Medicine (certificate no.: 4307272111102509853). Rats were maintained in the standard Laboratory Animal Center of Hunan University of Chinese Medicine (temperature 24–26°C, humidity 50–70%), given one week to adapt to the facility before the experiment.

Among the 52 rats, 12 rats were used as sham group, and the others were used as the model group. The model group's rats were subjected to T10 spinal cord injury using Hassan Shaker spinal cord transection [13]. 24 rats met the model standard which were randomly divided into the DBND group (n = 12) and the DBND+EA group (n = 12).

### 2.2 Modeling

After 7 days of routine feeding, the model group rats were fasted and given water for 24 h. The rats were anesthetized by intraperitoneal injection of 3% sodium pentobarbital at a dose of 50mg /kg (Merck KGaA, USA), fixed the rat in the supine position and positioned the 13th thoracic vertebra as a bony marker. Rats' subcutaneous tissue at the 8th to 9th thoracic vertebrae were dissected longitudinally, Separating the musculature on both sides of the spine, removing T8 and T9 spinous processes and vertebral arches, exposing the spinal cord, cutting off the spinal cord with NO. 11 surgical blade, the muscles were sutured, sterilizing the incision and its surroundings with 5% compound iodine, and suturing the skin.

In the sham group, the skin above the 8th and 9th thoracic vertebrae was cut to expose the spinal cord, but not cut. We used the Crede maneuver to assist voiding in the morning (at 06:00), in the afternoon (at 14:00), and in the evening (at 22:00) every day.

### 2.3 Animal model evaluation

The T10 spinal cord injury model was evaluated by observing the hindlimb motor and bladder voiding function. When the hindlimb of the rat was completely dragged while walking and the Basso, Beattie and Bresnahan (BBB) score was 0, the SSCI model was successful [14].

In addition, the rats were still unable to urinate spontaneously and retain urine after the bladder shock phase. The urethral orifice offered resistance to the flow of urine during manual urine expression. If these conditions were met, the model generation was considered successful.

### 2.4 EA intervention

DBND+EA were treated with EA (20 minutes per session, once daily for 10 consecutive days)

after spinal shock (19 days after model generation). The dimensions of the acupuncture needles were 0.30 mm×25 mm (Huatuo brand, China). The Ciliao (BL32), Zhongji (RN3) and Sanyinjiao (SP6) acupoints were chosen according to the 9th edition of *Experimental Acupuncturology and Government Channel and Points Standard GB12346-90 of China*. The needle insertion depths at these three points were 10 mm, 5 mm, and 5 mm, respectively. Anesthesia was induced by inhalation of isoflurane gas; the rats were then fixed on an in-house rat board and received EA. The following parameters were used for EA stimulation: density wave, a frequency of 10/50 Hz, and an intensity that caused limb tremor while remaining tolerable.

**2.5 Urodynamics tests**

After EA intervention for one week, all rats were anesthetized pentobarbital. Then, the rats were fixed in the supine position with no urine in the bladder. In this study, a urinary catheter was inserted into the bladder and 0.9% normal saline (25~35°C) was infused intravenously (100 μl/min). Finally, bladder pressure curve changes, maximum cystometric capacity (MCC) and leak point pressure (LPP) were recorded using MP150 multi-channel physiological recorder (Biopac, USA).

**2.6 Hematoxylin and eosin staining**

After anesthesia, normal saline was rapidly injected through the left ventricle for perfusion until the outflowing fluid became clear, after which 4% paraformaldehyde was injected into the ventricle for perfusion fixation. After 48 hours of fixation with 4% paraformaldehyde, the detrusor and bladder neck were embedded, sectioned, stained, sealed with neutral gum. Finally, the histomorphology of detrusor muscle and bladder neck was observed by light microscope.

**2.7 TMT quantitative proteomics analysis**

**2.7.1 TMT labeling**

Our TMT labeling and proteomic analyses were performed by Kangcheng Biotech, Inc (Guangzhou，China). The detrusor muscle and bladder neck of three rats from each group were dissected Completely and centrifuged, and the supernatant was transferred to the EP tube. 100 mg of supernatant from each the detrusor muscle and bladder neck of three rats from each group were dissected thoroughly and centrifuged, and the supernatant was transferred to the EP tube. sample were taken for reduction, alkylation, acetone precipitation and protein redissolution to obtain the corresponding polypeptides. Samples were then labeled with 20 μL of TMT10-plex isotope (Thermo Scientific, USA), mixed and centrifuged. Samples were analyzed by liquid chromatography-tandem mass spectrometry (LC-MS /MS) analysis.

**2.7.2 Proteomic analysis**

MaxQuant (Version 1.6.1.0) (Thermo Scientific, USA) was used for database searches and TMT-based quantitative analysis of the raw data obtained by LC–MS/MS analysis. The protein database was uniprot_RATTus_20190711_ISO. The false discovery rate (FDR) levels for polypeptides and proteins were controlled at 0.01. Finally, we normalized the 10 samples to make the total protein or median consistent for each group. Proteins with P<0.05, unique peptides ≥ 2 and fold change (FC)>1.2 or <1/1.2 were defined as differentially expressed proteins (DEPs).

**2.7.3 Bioinformatics analysis of DEPs**

KOBAS3.0 (http://kobas.cbi.pku.edu.cn/) was used to analyze Kyoto Encyclopedia of Genes and Genomes (KEGG) pathways (*Rattus norvegicus*, P < 0.05). Finally, the enrichment results were sorted according to the input number, and the first 10 or 15 KEGG pathways were screened, using STRING 11.5 (https://string-db.org/) and Cytoscape 3.9.1 to constructing protein–protein interaction (PPI) analysis network.

**2.8 WB protocol**

Detrusor and bladder neck tissues were lysed and centrifuged before determination with the BCA quantification kit (Thermo Scientific, USA). A certain amount of proproteinogen was thoroughly mixed with 4× loading buffer and then denatured in a boiling water bath. Protein samples were separated by 8% or 10% SDS–PAGE and were transferred to 0.45 μm polyvinylidene fluoride (PVDF) membranes. Then, the PVDF membranes were soaked in 5% skim milk from powder and cleaned. Membranes were incubated with rabbit anti-Col4a2 (1:1000, Abclonal, Wuhan, China), rabbit anti-Gnaz (1:1000, Abclonal, Wuhan, China), rabbit anti-Kcnmb1 (1:1000, Thermo, Massachusetts, USA), rabbit anti-Smtn (1:1000, Abclonal, Wuhan, China) and mouse anti-β-actin (1:8000, Abcam, Cambridge, UK) overnight at 4 °C. Goat anti-rabbit IgG secondary antibody (1:10000, Elabscience, Wuhan, China) and Goat anti-mouse IgG secondary antibody (1:10000, Elabscience, Wuhan, China) was added. Finally, the PVDF membrane was reacted with freshly prepared enhanced chemiluminescence (ECL) solution (Advansta, USA) for 2 min and then rapidly exposed and developed in the dark.

**2.9 Statistical analysis**

SPSS 26.0 statistical software was used for analysis, data expressed as the mean ± standard deviation (X ± S). One-way analysis of variance (ANOVA) was used for comparisons between groups, T test used to pairwise comparisons. $P<0.05$ was considered statistically significant.

## 3. Results

**3.1 EA can improve urinary function in DBND after SSCI**

The rats in the sham group were in good general condition. In DBND and DBND+EA, all voluntary hind limb movement disappeared after spinal shock, and the rats moved by dragging themselves. The bladder was enlarged in the lower abdomen when palpated. The lower abdomens and cage lining pads of the rats were slightly damp, and resistance was felt when manual urination was performed. After EA intervention, the degree of bladder distension and the degree of resistance to manually assisted urination in the DNBD+EA were reduced. Figure 1A shows that the daily manually assisted urine output in the DBND fluctuated greatly after spinal cord shock, while the DBND+EA showed a slight downward trend.

Comparing with sham group，Urodynamics showed that LPP and MCC in DBND were obviously increased ($P<0.01$). LPP and MCC in the DBND+EA were significantly decreased ($P<0.01$, $P<0.05$) after EA intervention (Figure 1B and C).

HE staining showed that the detrusor mucosa of DBND had focal shedding of the mucosal epithelium (arrow ①), enlarged smooth muscle nuclei (arrow ②), muscle fiber hypertrophy, smooth muscle wall thickening, and collagen fiber hyperplasia in the muscle layer (arrow ③). EA treatment markedly reduced these anomalies (Figure 1D). In addition, a mass of fibroblasts (arrow ④) and inflammatory cells (arrow ⑤) were infiltrated in the lamina propria of bladder neck of DBND rats, smooth muscle cell nuclei were enlarged (arrow ②), and muscle fibers were thickened. EA significantly improved the pathologic changes of the bladder neck (Figure 1E).

These results indicate that EA can improve the urinary ability of DBND after SSCI.

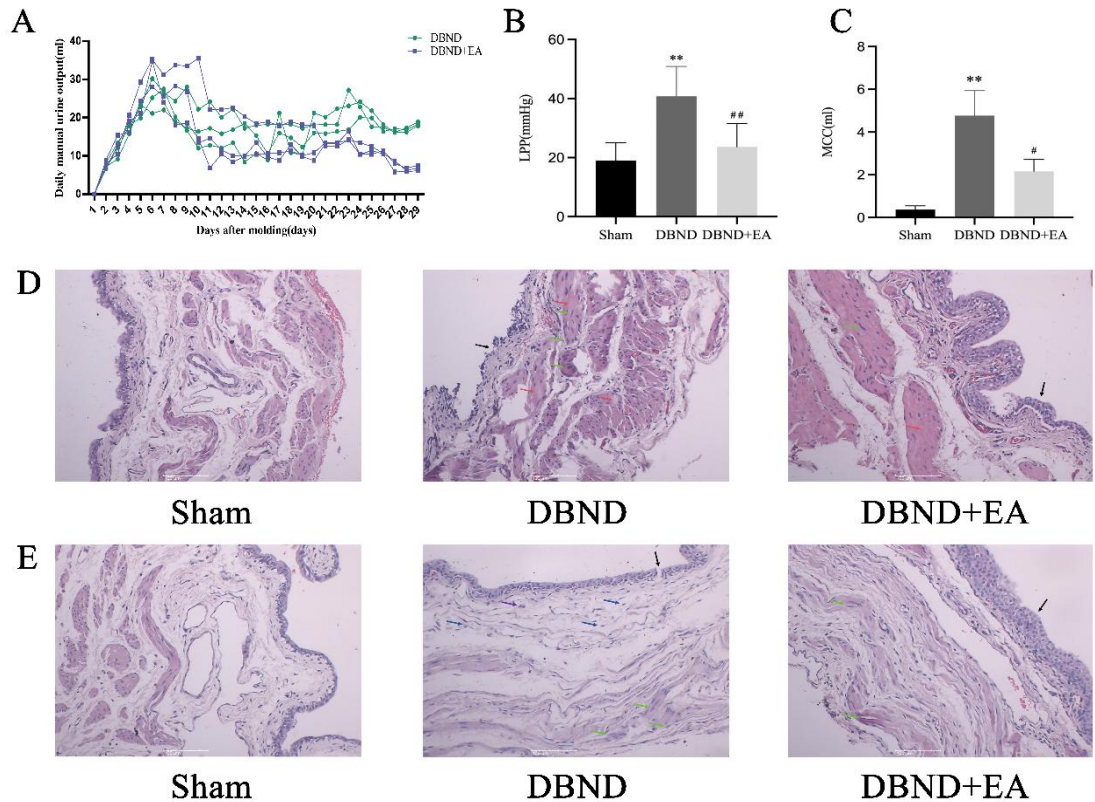

**Figure 1.** Therapeutic effect of EA for DBND. (A) Line chart of daily manual-assisted urine output of rats in DBND and DBND+EA. (B) Comparison of leak point pressures (LPP) among the three groups. (C) Comparison of Maximum cystometric capacity (MCC) among the three groups. *Compared with sham group, $P < 0.05$; **Compared with sham group, $P < 0.01$; # Compared with DBND, $P < 0.05$; ## Compared with DBND, $P < 0.01$. (D) Morphological changes of bladder detrusor in different groups under a light microscope after hematoxylin and eosin stain (×200). (E) Morphological changes of bladder neck in different groups under a light microscope after hematoxylin and eosin stain (×200). Black arrow: focal shedding of the mucosal epithelium; Green arrow: enlarged smooth muscle nuclei; Red arrow: collagen fiber hyperplasia; Blue arrow: fibroblasts; Purple arrow: inflammatory cells.

## 3.2 TMT quantitative proteomic analysis of the detrusor

### 3.2.1 TMT analysis of DEPs

Detrusor TMT results showed that a total of 4470 quantifiable proteins were detected. 230 DEPs (126 upregulated, 104 downregulated) were present in Sham/DBND (Figures 2A); 172 DEPs (93 upregulated and 79 downregulated) were present in DBND+EA/DBND (Figures 2B). 30 overlapping DEPs were found in Sham/DBND and DBND+EA/DBND (Fig. 2C). In the DBND group, 29 overlapping DEPs were reversed by EA and 1 overlapping DEPs was consistent with DBND (Table 1).

**Table 1.** 30 DEPs of bladder detrusor regulated by EA.

| Majority protein ID | Gene names | Protein description | Unique peptides | Score | Sequence coverage | FC (DBND/sham) | FC (DBND+EA/DBND) |
|---|---|---|---|---|---|---|---|
| F1M3W5 | Dmxl2 | Dmx-like 2 | 33 | 102.06 | 11.3 | 0.80 | 1.29 |
| P20760 | Igg-2a | Ig gamma-2A chain C region | 6 | 106.22 | 29.2 | 0.40 | 1.56 |
| H9KVF6 | Stk10 | Non-specific serine/threonine protein kinase | 10 | 44.55 | 9.7 | 1.35 | 0.72 |
| A0A0G2K946 | Spock2 | SPARC/osteonectin, cwcv and kazal-like domains proteoglycan 2 | 5 | 26.65 | 10.6 | 0.69 | 1.29 |
| P62255 | Ube2g1 | Ubiquitin-conjugating enzyme E2 G1 | 3 | 7.71 | 17.1 | 1.36 | 0.81 |
| P55770 | Nhp2l1 | NHP2-like protein 1 | 4 | 67.52 | 39.1 | 1.25 | 0.75 |
| Q5PPN5 | Tppp3 | Tubulin polymerization-promoting protein family member 3 | 12 | 247.22 | 46 | 0.64 | 1.29 |
| P35465 | Pak1 | Serine/threonine-protein kinase PAK 1 | 6 | 86.55 | 23.5 | 1.38 | 0.75 |
| P50411 | Ppp1r2 | Protein phosphatase inhibitor 2 | 7 | 22.12 | 27.3 | 1.36 | 0.82 |
| Q6PDU7 | Atp5l | ATP synthase subunit g, mitochondrial | 4 | 46.71 | 32 | 0.50 | 1.46 |
| G3V836 | Clu | Clusterin | 11 | 89.94 | 25.5 | 0.62 | 1.43 |
| F1LST4 | Mapt | Microtubule-associated protein | 9 | 199.21 | 27 | 0.65 | 1.28 |
| P19627 | Gnaz | Guanine nucleotide-binding protein G(z) subunit alpha | 5 | 31.01 | 17.7 | 0.61 | 1.32 |
| B1WBY5 | Dnajc11 | DnaJ (Hsp40) homolog, subfamily C, member 11 | 9 | 17.13 | 13.2 | 0.74 | 1.27 |
| A0A0H2UHF9 | Ablim2 | Actin-binding LIM protein 2 | 8 | 20.27 | 13.1 | 0.83 | 1.27 |
| P34901 | Sdc4 | Syndecan-4 | 3 | 3.81 | 13.9 | 1.28 | 1.25 |
| D3ZF44 | LOC684499 | Similar to stefin A2 | 2 | 1.69 | 15.5 | 2.14 | 0.40 |
| O35077 | Gpd1 | Glycerol-3-phosphate dehydrogenase [NAD (+)], cytoplasmic | 15 | 116.02 | 37.5 | 0.63 | 1.38 |
| P14046 | A1i3 | Alpha-1-inhibitor 3 | 10 | 323.31 | 23.2 | 0.51 | 1.48 |
| P02770 | Alb | Albumin | 37 | 323.31 | 55.4 | 0.69 | 1.39 |
| Q00238 | Icam1 | Intercellular adhesion molecule 1 | 4 | 6.50 | 7 | 1.56 | 0.64 |
| P37377 | Snca | Alpha-synuclein | 5 | 166.57 | 57.1 | 0.71 | 1.24 |
| A0A1B0GWV6 | Aox3 | Aldehyde oxidase 3 | 6 | 5.34 | 7.4 | 0.64 | 1.30 |
| Q5XIH1 | Aspn | Asporin | 9 | 40.28 | 17.6 | 0.70 | 1.37 |
| G3V733 | Syn2 | Synapsin-2 | 8 | 143.82 | 19.6 | 0.73 | 1.21 |
| A0A0G2JVW1 | Igfbp5 | Insulin-like growth factor-binding protein 5 | 5 | 8.84 | 16.2 | 0.73 | 1.32 |
| F1M6Q3 | Col4a2 | Collagen type IV alpha 2 chain | 7 | 78.88 | 5.4 | 1.70 | 0.70 |
| D3ZTP0 | Aldh1l2 | 10-formyltetrahydrofolate dehydrogenase | 9 | 38.51 | 13.9 | 1.29 | 0.80 |
| P47967 | Lgals5 | Galectin-5 | 5 | 92.92 | 39.3 | 0.44 | 3.98 |
| A0A0G2K4G5 | Fndc1 | Fibronectin type III domain-containing protein 1 | 3 | 6.11 | 1.8 | 1.32 | 0.73 |

### 3.2.2 KEGG pathway analysis of DEPs

Based on the 30 DEPs of the detrusor, 10 KEGG pathways were screened out ($P < 0.05$). KEGG pathway analysis found that the significantly enriched pathways included the MAPK pathway, focal adhesion, and ECM-receptor interaction, etc (Fig. 2D).

### 3.2.3 PPI analysis of DEPs

The interactions of 30 proteins are shown in the PPI. Each node in the PPI network represents a DEP The darker the color of the node, the more edges connected to the DEP (Fig. 2E).

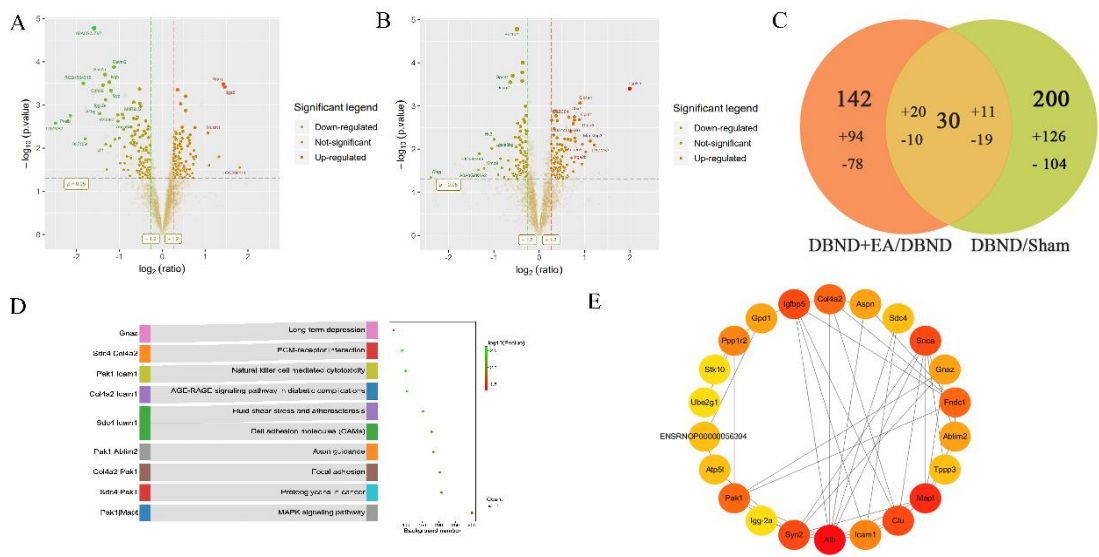

**Figure 2.** TMT quantitative proteomic analysis of bladder detrusor. (A) Volcano map of DBND/sham with 230 DEPs. (B) Volcano map of DBND+EA/ DBND with 172 DEPs. (C) Venny plots of DEPs and 30 overlapping proteins of DBND/sham and DBND+EA/DBND."+" represented up-regulated DEPs and "-" represented down-regulated DEPs. (D) KEGG pathway map of 30 DEPs. The abscissa was the background value of KEGG pathway enrichment, the ordinate is the $P$ value of pathway enrichment, the point size is the number of genes, and the left

side is the name of the gene involved in the pathway. (E) PPI plot of 30 DEPs.

### 3.3 TMT quantitative proteomic analysis of the bladder neck

#### 3.3.1 TMT analysis of DEPs

The results of TMT showed that 4049 quantifiable proteins were detected in bladder neck. 614 DEPs (306 upregulated, 308 downregulated) were present in Sham/DBND (Figures 3A); 145 DEPs (56 upregulated and 79 downregulated) were present in DBND+EA/DBND (Figures 3B). 59 overlapping DEPs were found in the Sham/DBND and DBND+EA/DBND (Fig. 3C). The expression trends of these DEPs in the DBND were reversed after EA (Table 2).

**Table 2.** 59 DEPs of bladder neck regulated by EA.

| Majority protein ID | Gene names | Protein description | Unique peptides | Score | Sequence coverage | FC (DBND/sham) | FC (DBND+EA/DBND) |
|---|---|---|---|---|---|---|---|
| Q68FR2 | Bin2 | Bridging integrator 2 | 3 | 8.62 | 8.1 | 2.56 | 0.47 |
| D4ABA5 | Smtn | Smtn protein | 34 | 323.31 | 34.2 | 0.42 | 1.87 |
| P25977-2 | Ubtf | Nucleolar transcription factor 1 | 2 | 4.18 | 2.8 | 0.61 | 1.40 |
| Q6AXZ0 | Paip2 | Polyadenylate-binding protein-interacting protein 2 | 2 | 2.13 | 12.1 | 1.30 | 0.78 |
| O55159 | Epcam | Epithelial cell adhesion molecule | 5 | 5.94 | 8.9 | 4.40 | 0.61 |
| Q6IRF7 | Cx3cl1 | Chemokine (C-X3-C motif) ligand 1 | 2 | 2.03 | 5.3 | 0.29 | 1.74 |
| A0A0G2K4L3 | Zfp622 | Zinc finger protein 622 | 2 | 3.54 | 2.5 | 0.72 | 1.25 |
| A0A0G2K6P1 | Hsd17b8 | (3R)-3-hydroxyacyl-CoA dehydrogenase | 4 | 20.58 | 21.1 | 0.69 | 1.35 |
| P17988 | Sult1a1 | Sulfotransferase 1A1 | 5 | 10.36 | 17.2 | 0.60 | 1.42 |
| P02634 | S100g | Protein S100-G | 2 | 3.92 | 24.1 | 2.78 | 0.25 |
| D4AC89 | Btbd11 | BTB domain-containing 11 | 5 | 7.72 | 4.3 | 0.58 | 1.33 |
| A0A0G2K6I3 | Nexn | Nexilin | 11 | 19.02 | 11.6 | 0.47 | 1.65 |
| D3ZLC1 | Lmnb2 | Lamin B2 | 21 | 140.78 | 35.2 | 0.73 | 1.25 |
| P63029 | Tpt1 | Translationally-controlled tumor protein | 5 | 18.92 | 18 | 1.84 | 0.74 |
| Q99JE6 | Plcb3 | 1-phosphatidylinositol 4,5-bisphosphate phosphodiesterase beta-3 | 14 | 27.34 | 10.6 | 0.73 | 1.22 |
| A0A0H2UI26 | Steap3 | Metalloreductase STEAP3 | 7 | 20.60 | 14.5 | 1.41 | 0.70 |
| Q9JKS6-2 | Pclo | Protein piccolo | 19 | 46.44 | 4 | 0.33 | 2.38 |
| Q5XFX0 | Tagln2 | Transgelin-2 | 9 | 186.01 | 53.8 | 1.65 | 0.76 |
| P0C5H9 | Manf | Mesencephalic astrocyte-derived neurotrophic factor | 5 | 55.47 | 20.7 | 1.22 | 0.74 |
| F1M943 | Armc8 | Armadillo repeat-containing protein 8 | 5 | 9.47 | 7.1 | 0.65 | 1.40 |
| Q66H12 | Naga | Alpha-N-acetylgalactosaminidase | 6 | 11.93 | 12 | 1.63 | 0.76 |
| P97678 | Kcnmb1 | Calcium-activated potassium channel subunit beta-1 | 3 | 9.07 | 17.3 | 0.67 | 1.31 |
| F1LUN5 | Col4a5 | Collagen type IV alpha 5 chain | 2 | 1.34 | 2.5 | 0.58 | 1.23 |
| Q66HD3 | Nasp | Nuclear autoantigenic sperm protein | 5 | 9.35 | 5.5 | 1.36 | 0.65 |
| A0A140TAC3 | Epn1 | Epsin-1 | 3 | 16.90 | 6.6 | 1.47 | 0.73 |
| Q5XIG5 | Gkap1 | G kinase-anchoring protein 1 | 3 | 5.62 | 6.8 | 0.78 | 1.31 |
| Q499V1 | Upp1 | Uridine phosphorylase | 7 | 32.83 | 17.9 | 1.97 | 0.59 |
| B2RZ44 | Naa20 | N(alpha)-acetyltransferase 20, NatB catalytic subunit | 3 | 2.33 | 14.4 | 0.45 | 1.90 |
| Q2KN99 | Specc1l | Cytospin-A | 6 | 5.08 | 4.7 | 0.61 | 1.37 |
| A0A0G2K737 | Txnl1 | Thioredoxin-like protein 1 | 9 | 154.08 | 40.4 | 1.62 | 0.80 |
| Q9WVH8 | Fbln5 | Fibulin-5 | 7 | 76.60 | 14.7 | 0.42 | 1.28 |
| Q6Q0N1 | Cndp2 | Cytosolic non-specific dipeptidase | 11 | 123.98 | 23.8 | 1.40 | 0.72 |
| F1M3G7 | Akap13 | A-kinase anchor protein 13 | 5 | 9.46 | 1.7 | 1.50 | 0.77 |
| Q9R066-2 | Cxadr | Coxsackievirus and adenovirus receptor homolog | 3 | 3.15 | 9.1 | 1.65 | 0.70 |
| Q5U2R7 | Mesdc2 | LRP chaperone MESD | 6 | 11.85 | 23.2 | 1.28 | 0.65 |
| F1LTF8 | Lama4 | Laminin subunit alpha 4 | 37 | 316.31 | 18.4 | 0.40 | 1.33 |
| P62914 | Rpl11 | 60S ribosomal protein L11 | 8 | 56.21 | 36.5 | 1.30 | 0.81 |
| F1LM84 | Nid1 | Nidogen-1 | 29 | 265.56 | 24.2 | 0.38 | 1.33 |
| Q3ZBA0 | Tecpr1 | Tectonin beta-propeller repeat-containing protein 1 | 3 | 2.84 | 2.5 | 0.35 | 1.65 |
| Q4KLH7 | Rad21 | RAD21 cohesin complex component | 5 | 14.83 | 7.9 | 0.77 | 1.22 |
| D3ZAF7 | Tbc1d2b | TBC1 domain family, member 2B | 7 | 11.29 | 7.4 | 1.29 | 0.70 |
| D3ZUM4 | Glb1 | Beta-galactosidase | 8 | 105.77 | 15.1 | 1.53 | 0.71 |
| Q66H91 | Git2 | ARF GTPase-activating protein GIT2 | 4 | 3.93 | 6.2 | 1.30 | 0.78 |
| Q6AYS8 | Hsd17b11 | Estradiol 17-beta-dehydrogenase 11 | 8 | 19.13 | 24.2 | 1.34 | 0.79 |
| A0JPP1 | Drap1 | Dr1-associated corepressor | 2 | 11.51 | 10.2 | 0.80 | 1.27 |
| Q6IN22 | Ctsb | Cathepsin B | 8 | 168.77 | 21.5 | 1.89 | 0.60 |
| A0A0G2K2V4 | C2cd2l | C2CD2-like | 5 | 6.06 | 7.2 | 0.58 | 1.37 |
| F1LS67 | Rgs6 | Regulator of G-protein-signaling 6 | 3 | 23.78 | 14.3 | 0.55 | 1.45 |
| Q4FZU4 | Adamtsl4 | ADAMTS-like protein 4 | 2 | 4.87 | 1.9 | 0.68 | 1.35 |
| Q642A6 | Vwa1 | von Willebrand factor A domain-containing protein 1 | 6 | 31.45 | 14.5 | 0.29 | 1.48 |
| D4A8N1 | Dpm1 | Dolichol-phosphate mannosyltransferase subunit 1 | 5 | 11.33 | 18.5 | 1.38 | 0.81 |
| D3ZGW2 | Ap1g2 | AP-1 complex subunit gamma | 3 | 5.49 | 3.1 | 1.71 | 0.74 |
| D3ZC49 | Stfa2l3 | Stefin A2 | 4 | 3.90 | 35.9 | 2.42 | 0.38 |
| Q68FT3 | Pyroxd2 | Pyridine nucleotide-disulfide oxidoreductase domain-containing protein 2 | 4 | 16.58 | 7.4 | 0.73 | 1.29 |
| G3V6T7 | Pdia4 | Protein disulfide-isomerase A4 | 26 | 287.24 | 34.2 | 1.25 | 0.78 |
| M0RCH5 | Gnpda1 | Glucosamine-6-phosphate isomerase | 4 | 30.42 | 19.2 | 1.41 | 0.72 |
| P03994 | Hapln1 | Hyaluronan and proteoglycan link protein 1 | 11 | 83.86 | 29.1 | 0.63 | 1.22 |
| B2GUZ9 | Fam49b | Fam49b protein | 8 | 45.60 | 29 | 1.23 | 0.80 |
| Q9JKB7 | Gda | Guanine deaminase | 22 | 323.31 | 42.3 | 1.43 | 0.71 |

### 3.3.2 KEGG pathway analysis of DEPs

KEGG enrichment analysis was performed on 59 DEPs in the bladder neck, and it was found that 15 were significantly enriched in the cGMP-PKG pathway, vascular smooth muscle contraction, the Relaxin pathway, etc (Fig. 3D).

### 3.3.3 PPI analysis of DEPs

The interactions of 59 proteins are shown in the PPI. Each node in the PPI network represents a differentially expressed protein. The darker the color of the node, the more edges connected to the DEPs (Fig. 3E).

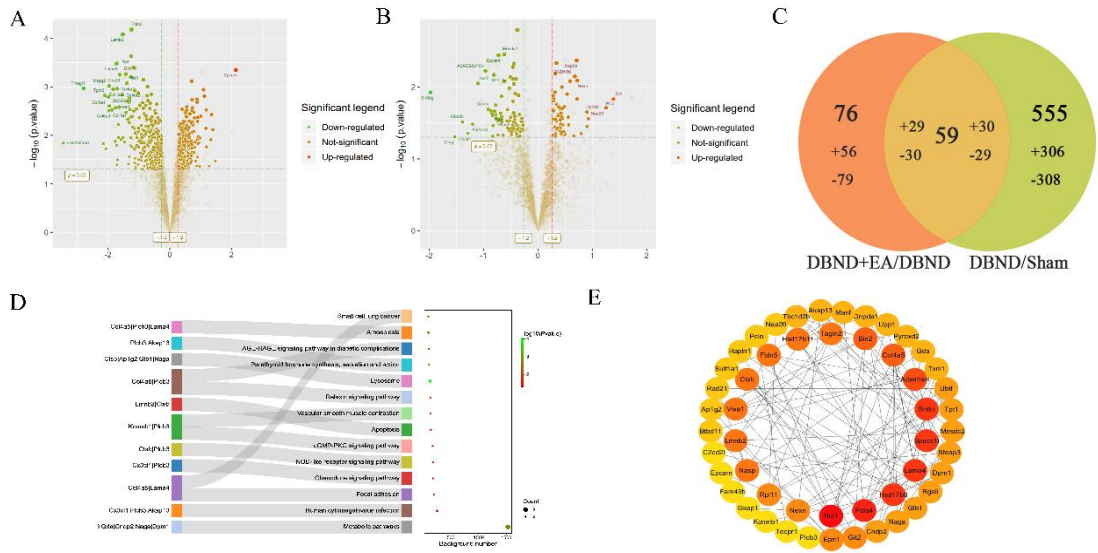

**Figure 3.** TMT quantitative proteomic analysis of bladder neck. (A) Volcano map of DBND/sham with 614 DEPs. (B) Volcano map of DBND+EA/ DBND with 145 DEPs. (C) Venny plots of DEPs and 59 overlapping proteins of DBND/sham and DBND+EA/DBND."+" represented up-regulated DEPs and "-" represented down-regulated DEPs. (D) KEGG pathway map of 59 DEPs. The abscissa was the background value of KEGG pathway enrichment, the ordinate is the $P$ value of pathway enrichment, the point size is the number of genes, and the left side is the name of the gene involved in the pathway. (E) PPI plot of 59 DEPs.

### 3.4 WB confirmation of TMT-based results

We respectively selected 2 DEPs related to bladder function from the TMT results of the detrusor and bladder neck for WB verification. The WB results of Col4a2 and Gnaz in the detrusor and Smtn and Kcnmb1 in the bladder neck were consistent with the TMT results (Fig. 4).

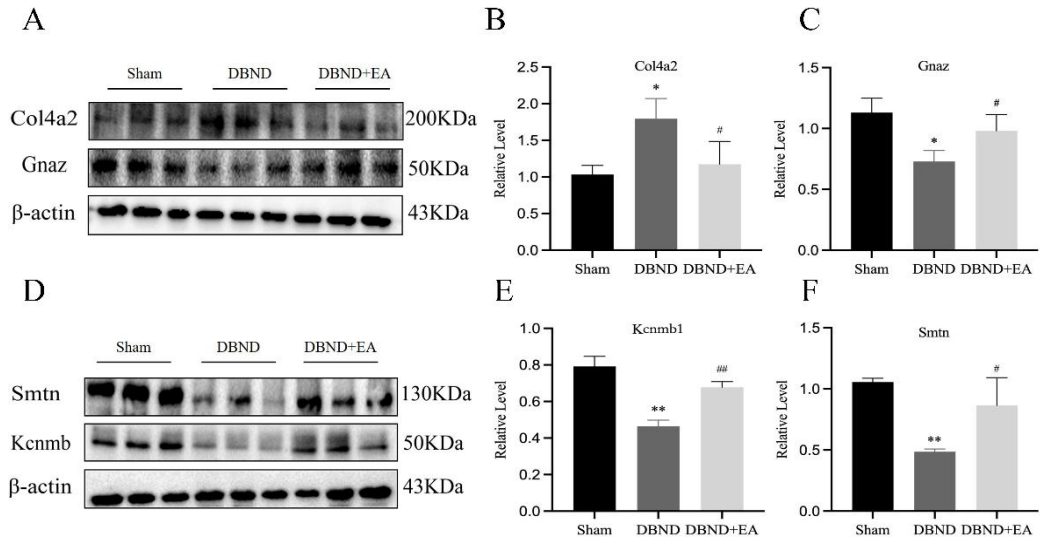

**Figure 4.** Validation of TMT quantitative proteomic analysis results by WB. The WB results were consistent with the TMT quantitative proteomic analysis results. *$P < 0.05$, compared with sham group; **$P < 0.01$, compared with sham group; # $P < 0.05$, compared with DBND; ## $P < 0.01$, compared with DBND.

## 4. Discussion

Neurogenic bladder after SCI has always been challenging to address with clinical intervention. According to the pathological characteristics of DBND, the detrusor and bladder neck are simultaneously used as research targets, and it has positive clinical significance in that it promotes the effective contraction of detrusor and inhibita spinal cord. Ciliao (BL32) is located in the second posterior sacral foramen, below the musc overactivity of the bladder neck. There are many records of acupuncture treating this disease in ancient Chinese medical books, and modern clinical researches have also verified that EA has a good curative effect on neurogenic bladder. Acupuncture points can stimulate the vegetative nerves of corresponding spinal cord segments, thereby regulating visceral activity and exerting bidirectional, integrated multitarget regulatory effects. The Zhongji (RN3) acupoint area is dominated by T12-L1 spinal cord. Acupuncture at Zhongji (RN3) can adjust the T12-L1 sympathetic nerve area. The S2 segment of the second sacral nerve innervates the Sanyinjiao (SP6) area. parasympathetic activity of the second sacral nerve and S2-S4 could be elicited by acupuncture at Sanyinjiao (SP6) and Ciliao (BL32) [15]. Therefore, acupuncture at Zhongji (RN3), Sanyinjiao (SP6) and Ciliao (BL32) can regulate the sympathetic and parasympathetic nerves that innervate bladder and simultaneously act on the detrusor and bladder neck, sequentially regulating bladder function.

In this study, we found that after EA treatment, detrusor edema was reduced, thickened smooth muscle fibers were decreased, and the proliferation of collagen fibers was reduced; the inflammatory response was reduced and the histological morphology was clear in the bladder neck. Besides, the MCC and LPP in the EA group were reduced. These data indicated that EA at Zhongji (RN3), Sanyinjiao (SP6) and Ciliao (BL32) can promote the effective contraction of the detrusor, reduce bladder outlet resistance and improve urinary function in model rats. This is consistent with previous studies [11].

To explore the potential targets and internal mechanisms of EA treatmenting SSCI-induced DBND, we used TMT quantitative proteomic analysis for proteomic analysis of detrusor and bladder neck tissue and screened these tissues for DEPs. The TMT results showed that after EA treatment, 30 DEPs in the detrusor muscle and 59 DEPs in the bladder neck were regulated in a manner that counteracted the expression changes associated with DBND.

Next, we performed a bioinformatic analysis of DEPs, and the results showed that focal adhesion, the MAPK pathway and ECM-receptor interaction, which were among the noteworthy KEGG pathways of 30 DEPs in the detrusor, were closely associated with smooth muscle cell contraction (Fig. 4). Additionally, the cGMP-PKG pathway, the relaxin pathway and vascular smooth muscle contraction, which were among the enriched KEGG pathways of the 59 DEPs in the bladder neck, were significantly correlated with tissue fibrosis and smooth muscle cell contraction.

We selected collagen type IV alpha 2 (Col4a2) and guanine nucleotide-binding protein G(z) subunit alpha (Gnaz) from 30 DEPs in detrusor for WB verification and found that EA could reduce Col4a2 and increase Gnaz in detrusor, which was in accordance with TMT results. Of the 59 DEPs in bladder neck, Smoothhelin (Smtn, contractile phenotype smooth muscle cells marker) and calcium-activated potassium channel subunit beta-1 (Kcnmb1, inducing smooth muscle cell membrane hyperplasia) were selected for WB validation. It was found that EA can upregulate the expression of Smtn and Kcnmb1 in the bladder neck. The results of TMT were consistent with those of WB.

Our study shows that the MAPK pathway was significantly enriched in the detrusor, and insulin-like growth factor-binding protein 5 (Igfbp5) was the critical DEP involved in MAPK pathway. The MAPK pathway is connected to cell proliferation and differentiation, playing a salient role in insulin growth factor (IGF)-mediated tissue fibrosis [16]. Studies have shown that the binding of IGF to its receptor can activate the RAS/ERK/c-fos pathway to induce proliferation of fibroblasts capable of synthesizing and secreting collagenous fiber, which ultimately leads to an increase in the extracellular matrix and fibrosis of tissues [17]. Fibrosis of muscle tissue affects the normal contractile function of this tissue, which is why relaxin is clinically used to treat muscle tissue fibrosis [18]. However, IGF must be dissociated from insulin-like growth factor binding protein (Igfbp) before binding to corresponding receptor and performing a biological function. As a member of the Igfbp family, Igfbp5 is an effective inhibitor of the biological effects of IGF. Secreted Igfbp5 interacts with an intracellular feedback mechanism to inhibit the activation of IGF signaling [19] (fig 5). Among the 30 DEPs screened in detrusor, Igfbp5 expression was observably higher in DBND+EA than in the DBND. In addition, the expression of collagen type IV alpha 2 (Col4a2) and fibronectin type III domain-containing protein 1 (Fndc1) in DBND+EA was dramatically decreased. We validated Col4a2 using WB and found that EA reduced the expression of Col4a2 in detrusor muscle, which was in keeping with the TMT data.

Studies have shown that Col4a2 and Fndc1 have an important connection in process of tissue fibrosis which is an important component of the extracellular matrix (ECM) [20]. In our experiment, the expression of Col4a2 and Fndc1 in the detrusor decreased after EA treatment, which indicates that EA treatment can reduce the proliferation and differentiation of fibroblasts in SSCI-induced neurogenic bladder, thereby reducing detrusor fibers and improving bladder function. Early studies have found that ECM-receptor interactions can affect cell adhesion and cell signaling and lead to changes in the cytoskeleton [21]. As an important part of the ECM, Col4a2 is involved in

ECM-receptor interaction signaling. In ECM-receptor interactions, Col4a2 binds to integrin receptors on detrusor and forms focal adhesions with cytoskeletal proteins, which can ultimately activate a series of downstream molecules. Col4a2 binds to integrins activates the focal adhesion kinase (FAK)–RAC–serine/threonine protein kinase 1 (PAK1) signaling pathway by activating in situ FAK [22]. PAK1 can inhibit the activity of myosin light-chain kinase (MLCK), thereby reducing the phosphorylation of myosin light chain (MLC) and ultimately affecting the contraction function of smooth muscle [23]. In this study, it was found that the expression of PAK1 in detrusor was decreased significantly after EA treatment compared with DBND, which promoted the activation of MLCK (fig 5). In this study, it was found that the expression of PAK1 in the detrusor decreased significantly after EA treatment compared with the DBND, which resulted in a weakening of the inhibitory effect on MLCK. The increase in MLCK activity in the DBND+EA further promoted MLC phosphorylation, which Suggested that EA may ameliorate the urination efficiency by decreasing degree of fibrosis and increasing the effective contraction of the detrusor.

Guanylate-binding proteins (G proteins) play an important status in the cell signal transduction pathway and are made up of three different subunits, α, β, and γ. Among them, Gα protein is further divided into Gs, Gi, Go, Gq, G12, and G13, and Gnaz belongs to the Gi family [24]. When the cellular receptor is coupled to Gi, it inhibits adenylate cyclase (AC) activation, which in turn inhibits downstream signaling pathway activated by cyclic adenosine monophosphate (cAMP) [25]. Studies have shown that cAMP can regulate smooth muscle cell contraction by activating the PKA/Pho/Rock/MLC pathway [26]. Therefore, the endogenous relaxant PTHrP, which affects bladder contraction, inhibits the spontaneous contraction of the detrusor by increasing intracellular cAMP expression. Increased Gnaz expression inhibits cAMP production, which promotes smooth muscle contraction by affecting its downstream signaling pathways (fig 5). We verified Gnaz by WB and found that Gnaz expression in the detrusor of the DBND+EA was reduced compared with DBND, which was consistent with the TMT data. Furthermore, the ATP synthase subunit Atp5l is a component of ATP synthase in the mitochondrial membrane, and it is involved in the process by which ATP synthase generates ATP on the mitochondrial membrane to provide energy to cells [27]. Among the 30 DEPs enriched in detrusor muscle, Atp5l was significantly enriched, and Atp5l expression was observably increased after EA intervene (fig 5). This suggests that EA may promote the effective contraction of the detrusor muscle and increase urination efficiency by inhibiting the generation of cAMP and providing the energy needed for detrusor muscle contraction.

Another of the 30 DEPs, Synapsin-2 (Syn2) is associated with vesicle trafficking and neurotransmitter release at synapses. Syn2 is mainly used in the study of nervous system diseases. It acts by binding to synaptic vesicle membranes. Phosphorylated Syn2 separates from synaptic vesicles to promote vesicle exocytosis, chemical synaptic signal transmission and neurotransmitter release [28] (fig 5). Our study found that Syn2 expression in the detrusor of the DBND+EA were increased comparing with the DBND, which may be the potential molecular mechanism by which EA treats DBND after suprasacral spinal cord injury.

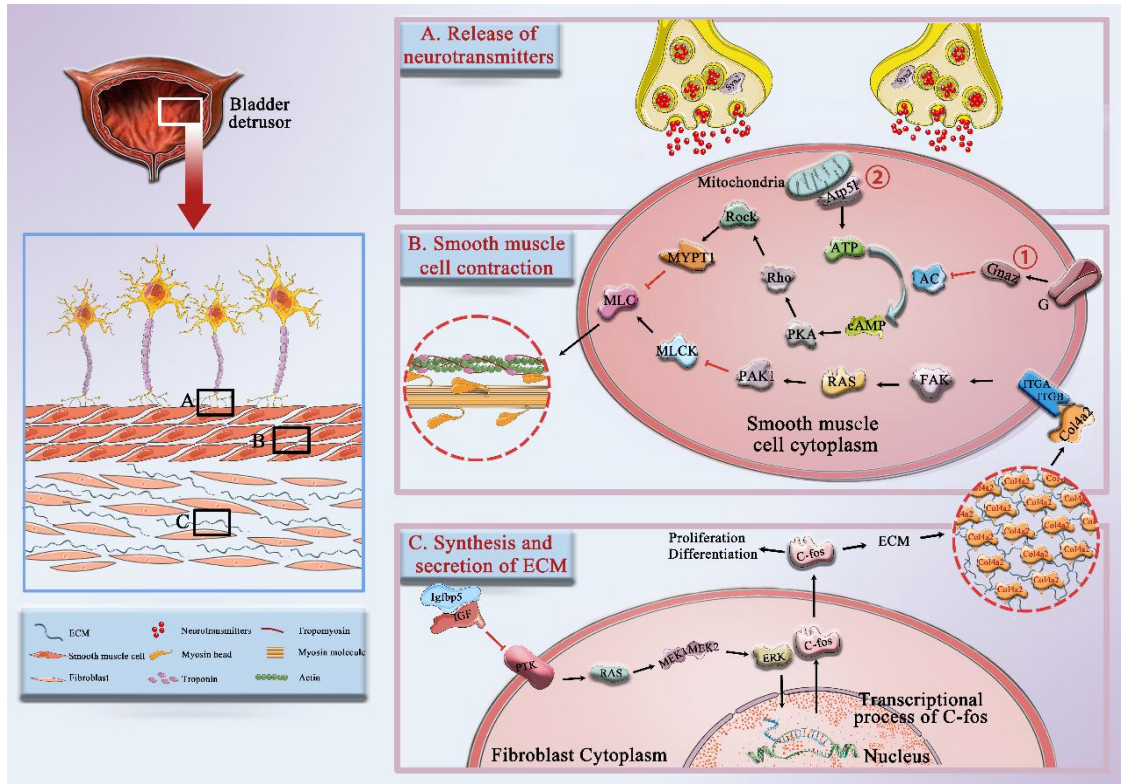

**Figure 5.** Potential association of EA-regulated 30 DEPs in bladder detrusor muscles. (A) Syn2 is involved in release of presynaptic membrane neurotransmitter release. (B) ① Gnaz affects the contraction of smooth muscle cells by regulating AC/cAMP/MYPT1/MLC signaling pathways. ② Atp5l affects smooth muscle contraction by affecting mitochondrial ATP synthesis. (C) Igfbp5 affects the synthesis and release of ECM through MAPK signaling pathway, further affecting bladder detrusor fibrosis. Col4a2 (a component of ECM) ultimately affects the contraction of detrusor through the ECM-receptor interaction pathway and Focal adhesion pathway.

Bioinformatics analysis of the bladder neck showed that the relaxin signaling pathway was significantly enriched. The relaxin signaling pathway is closely related to fibrosis in various tissues, and fibrosis of bladder tissue ultimately affects smooth muscle contraction and bladder tissue compliance [29]. Studies have shown that chronic inflammation is a common factor that stimulates tissue fibrosis [30]. According to the histomorphological results of this experiment, a mass of fibroblasts and inflammatory cells infiltrated the bladder neck of DBND. After EA treatment, the inflammatory response was improved. Smtn is a specific marker of mature contractile smooth muscle cells. When smooth muscle cells are induced to transform from contractile to synthetic due to various factors, Smtn is a hallmark protein that is first downregulated until it disappears [31]. When contractile smooth muscle cells are converted to synthetic smooth cells, the normal contractile function of the smooth muscle decreases, and extracellular matrix production increases, which may lead to further tissue fibrosis [32] (fig 6). After surgical modeling, the expression of Smtn in the bladder neck was significantly decreased, which indicated that the contractile smooth muscle cells decreased in the bladder neck after modeling. Compared with the DBND, the Smtn expression in bladder neck was obviously improved in EA-treated rats, which indicated that EA can maintain the stability of contractile smooth muscle cells and reduce the production of extracellular matrix and tissue fibrosis to maintain normal bladder function.

Furthermore, our results show that cGMP-PKG pathway and vascular smooth muscle contraction pathway were significantly enriched in bladder neck and are directly involved in smooth muscle relaxation and contraction [33]. Calcium-activated potassium channel subunit beta-1 (Kcnmb1) and 1-phosphatidylinositol 4,5-bisphosphate phosphodiesterase beta-3 (Plcb3) are the major DEPs involved in these two pathways. As an important regulatory subunit of Large-Conductance Calcium-Activated Potassium Channels (BKca), Kcnmb1 can increase the calcium sensitivity of BKca channels in the smooth cell membrane [34]. Activation of BKca channels enables efflux of intracellular $K^+$, which ultimately leads to smooth cell hyperplasia. Thus, Kcnmb1 can act as a negative regulator of smooth muscle contraction. Studies have shown that knocking out the Kcnmb1 gene in bladder smooth muscle reduces the activity of BKca channels, which ultimately affects the contraction of bladder smooth muscle [35]. The increased expression of Plcb3 releases $Ca^{2+}$ from the endoplasmic reticulum to the cytoplasm by activating the intracellular PLC/IP3/IP3R signaling pathway [36]. Increased intracytoplasmic $Ca^{2+}$ concentration activates BKca channels on the cell membrane, ultimately promoting the relaxation of contracted smooth muscle cells [37] (Fig 6). The TMT results showed that the expression of Kcnmb1 and Plcb3 in bladder neck after EA treatment was higher than that in DBND. In addition, the WB results showed that the Kcnmb1 expression in the bladder neck increased after EA treatment. In this way, EA may promote bladder neck relaxation by upregulating the expression of Kcnmb1 and Plcb3 in the bladder neck, thereby activating the cGMP-PKG signaling pathway.

Among the 59 DEPs screened in the bladder neck, we found that G protein signaling 6 (Rgs6) regulator was significantly increased and protein kinase A-anchored protein 13 (Akap13) was significantly lower after EA treatment compared with DBND. As a signal regulator on the cell membrane, Rgs6 can bind to the complex of GTP and Gi protein (Gnai) through the activity of GTPase accelerating proteins (GAPs), resulting in GTP hydrolysis, Gnai inactivation and termination of Gnai downstream signaling transduction [38]. In a physiological state, Gnai can inhibit AC activation and ultimately affect the cAMP pathway, and Activation of the downstream cAMP pathway is closely related to the contraction of smooth cells [39] (fig 6). The TMT results demonstrated that the expression of Rgs6 in DBND+EA was higher than that in the DBND, which could reduce the inhibitory effect of GNAI on the cAMP signaling pathway. A-kinase anchoring protein 13 (Akap13) is a Rho GTPase activating protein (Rho-GEF) in the cytoskeleton that can activate the Akap13/RhoA/Rock signaling system in cells, thereby inhibiting the phosphorylation of MLCP and MLC and finally affecting smooth muscle contraction [40] (fig 6). Akap13 expression was significantly decreased in the DBND+EA compared with the DBND, and ultimately affected the Akap13/RhoA/Rock/MLCP/MLC pathway. In conclusion, Rgs6 and Akap13 may be potential therapeutic targets for EA treating DBND.

Besides, the function of the bladder neck is not related only to signal transduction in smooth muscle cells; the neurotransmitters released by the nerve endings also control the relaxation and contraction of the bladder neck. The active zones (AZs) of the presynaptic membrane are the areas where synaptic vesicles bind to the presynaptic membrane to release neurotransmitters [41]. Among our enriched DEPs, Git2 and Pclo, as highly conserved proteins in AZs of the presynaptic membrane, participate in the presynaptic regulation of chemical synapses and affect the release of synaptic vesicles [42]. Studies have shown that Git2 can affect synaptic vesicle exocytosis by regulating actin cytoskeleton dynamics in AZs [43]. In addition, Pclo negatively regulates the

ubiquitination of proteins in AZs to protect synaptic cells and maintain synaptic integrity [44] (fig 6)

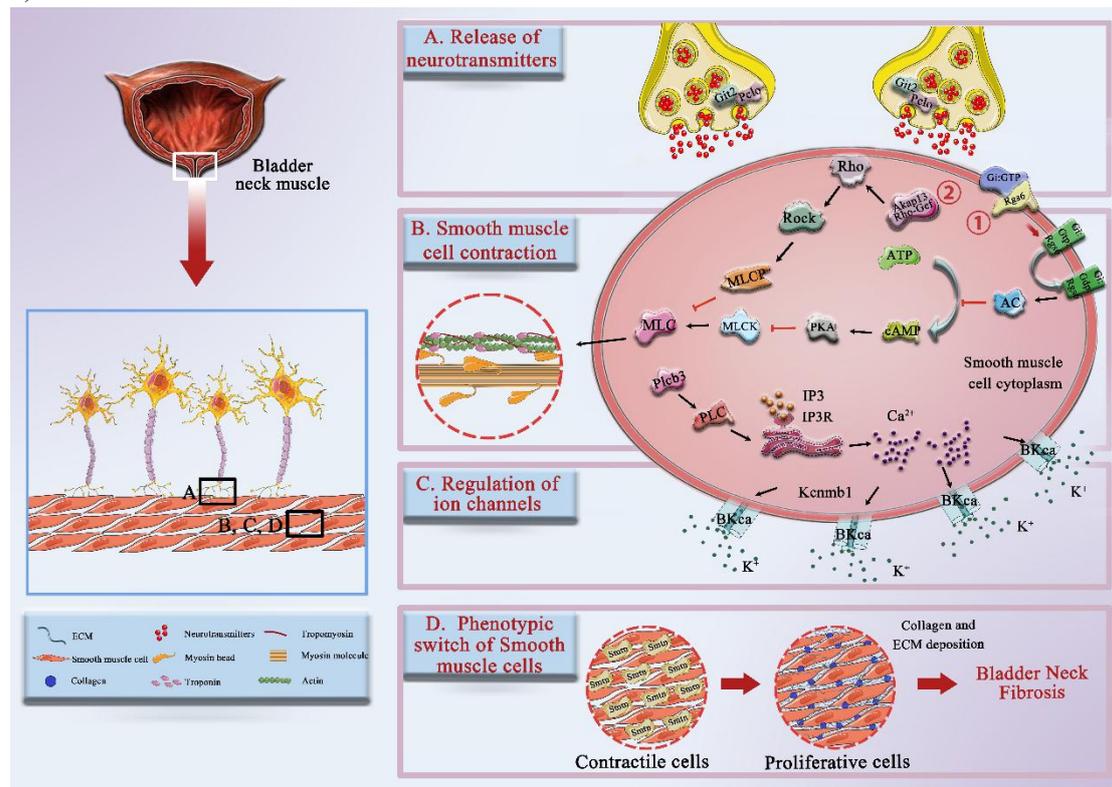

**Figure 6.** Potential association of EA-regulated 59 DEPs in bladder neck. (A) Git2 and Pclo is involved in neurotransmitters release from the AZs of the presynaptic membrane. (B) ① Rgs6 affects the contraction of smooth cells by regulating AC/cAMP/MLCK/MLC signaling pathways. ② Akap13 affects the contraction of smooth cells by regulating Rho/Rock/MLCP/MLC signaling pathways. (C) Plcb3 affects the opening of BKCa channel. (D) Phenotypic switch of smooth muscle cells affects bladder neck fibrosis.

In conclusion, we preliminarily believe that the possible mechanism of EA for DBND after SSCI is to improve the synergy of the detrusor and bladder neck by motivating detrusor contraction and bladder neck relaxation. Col4a2, Gnaz, Smtn, and Kcnmb1 were validated by WB in this study, but other DEPs presented in the discussion were not validated by WB. In our future research, we will verify each of the DEPs mentioned in the discussion. In addition, to further verify the functional mechanism of EA in the treatment of SSCI-induced DBND, we will carry out further research in which we use agonists and inhibitors to identify potential therapeutic targets.

## Data availability

The data have been uploaded to ProteomeXchange with identifier PXD034595 (http://proteomecentral.proteomexchange.org/cgi/GetDataset?ID=PXD034595).

## Ethics Approval and Consent to Participate

Experimental animal ethics committee of Hunan University of Chinese Medcine. Ethical approval number: LL2019092303.

## Author Contributions

Experiment design: KA, QL. Experiment development: LYT, QRQ, XW. Project management: KA, LQ. Analysis of data: LYT, JCL, YYL. Paper review: HZ. Paper writing: LYT, QRQ, KA. Experimental support: KA, QL. All authors approved the final version of the manuscript.

## Funding

We acknowledge the Natural Science Foundation of China (General) (No.81874510), Natural Science Foundation of Hunan Province (No.2022JJ40301), the Scientific research project of Hunan Provincial Department of Education National (No.21B0369) and Key scientific research projects of Hunan Administration of traditional Chinese Medicine (No. 201704).

## NOTES

The authors declare no competing financial interest.